\documentclass[fleqn,usenatbib]{mnras}

\usepackage{newtxtext,newtxmath}

\usepackage[T1]{fontenc}
\usepackage{ae,aecompl}

\usepackage{graphicx}	
\usepackage{amsmath}	
\usepackage{amssymb}	
\usepackage{color}      

\title[Slingshot prominences]{Slingshot prominences: nature's wind gauges}

\author[M. Jardine et al.]{
Moira Jardine,$^{1}$\thanks{E-mail: mmj@st-andrews.ac.uk}
Andrew Collier Cameron,$^{1}$
\\
$^{1}$SUPA, School of Physics and Astronomy, North Haugh, St Andrews, Fife, KY16 9SS, UK\\
}

\date{Accepted XXX. Received YYY; in original form ZZZ}

\pubyear{2018}

\begin{document}
\label{firstpage}
\pagerange{\pageref{firstpage}--\pageref{lastpage}}
\maketitle
\begin{abstract}
Mass loss rates for the tenuous, hot winds of cool stars are extremely difficult to measure, yet they are a crucial ingredient in the stars' rotational evolution. We present a new method for measuring these mass loss rates in young, rapidly-rotating  stars. These stars are known to support systems of  ``slingshot prominences''  fed by  hot wind material flowing up from the stellar surface into the summits of closed magnetic loop structures. The material gathers and cools near the co-rotation radius until its density becomes large enough that it is visible as a transient absorption feature in the hydrogen Balmer lines and strong resonance lines such as Ca II H \&\ K.  Here we present the key insight that the sonic point usually lies well below the condensation region. The flow at the wind base is therefore unaffected by the presence of an overlying prominence, so we can use the observed masses and recurrence times of the condensations to estimate the mass flux in the wind. These measurements extend the relationship between mass loss rate per unit surface area and X-ray flux to span 5 orders of magnitude. They demonstrate no evidence of the suspected weakening of stellar mass loss rates at high X-ray flux levels.
\end{abstract}




\section{Introduction}
The coronae of low mass stars (i.e. stars of mass M$_\star$ < 1M$_\odot$) are magnetically heated to temperatures of $10^6-10^7$ K. While much of this plasma is confined, some also escapes along magnetic field lines. This magnetic channelling ensures that this outflowing wind carries away significant angular momentum, even if the low density of these winds means that they remove little mass \citep{1958ApJ...128..677P, 1967ApJ...148..217W, 1968MNRAS.138..359M}. As a result, these winds govern the evolution of the star's rotation rate and hence its magnetic activity. They are, however, extremely difficult to observe. While the solar wind can be examined with in situ measurements, cool star winds are so tenuous that direct measurements are very challenging \citep{2004LRSP....1....2W}. 

Rotational evolution can however be used as an indirect method of testing stellar wind models. Distributions of rotation rates are now available for many open clusters of known ages, for example \citet{2009MNRAS.392.1456I}. These can be used to test rotational evolution models \citep{2013A&A...556A..36G,2015A&A...577A..98G} which are themselves based on scaling laws for angular momentum loss \citep{2011ApJ...741...54C,2012ApJ...745..101M}. Spin-down models can also be constructed to fit these distributions \citep{2015A&A...577A..27J,2015A&A...577A..28J}. This gives predictions for wind mass loss rates and velocities. One very interesting outcome of these many studies has been the very slow spin down of the lowest mass stars \citep{2012ApJ...746...43R} which raises questions about the role of the field geometry \citep{2015ApJ...807L...6G,2017MNRAS.465L..25J} and in particular the location of the ``source surface'' (\citet{2015ApJ...798..116R,2015ApJ...814...99R,2018MNRAS.474..536S}) which is the radius at which the stellar magnetic field becomes completely open. 

Of the direct methods of measuring stellar winds, one of the most straightforward comes from the thermal radio emission that an expanding wind produces \citep{1975A&A....39....1P}. This can provide measurements of the wind density, but this typically provides only upper limits through non-detections  \citep{1996ApJ...462L..91L,1997A&A...319..578V,2014ApJ...788..112V}. More recently, \citet{2017A&A...599A.127F} presented stringent upper limits on mass loss rates for four solar-type stars based on a range of optical depth regimes. These provide tests for predictions of the mass loss of the young Sun. Most other methods depend on using the interaction of the wind with some other body. Searches for the X-ray signature of charge exchange when the ionised wind interacts with the neutral interstellar medium have also provided upper limits \citep{2002ApJ...578..503W}. For stars in a binary system with a white dwarf, the pollution of the white dwarf photosphere from the wind of the companion may leave a detectable trace in the white dwarf spectrum. Modelling of this process provides estimates of the wind outflow rates \citep{2006ApJ...652..636D,2012MNRAS.420.3281P}. Most recently, the interaction of the escaping atmosphere of a planet with the stellar wind has provided a means of measuring the mass loss rate of the stellar wind  \citep{2016A&A...591A.121B,2017MNRAS.470.4026V,2014Sci...346..981K}.

One of the most promising techniques to date has been the detection of Lyman $\alpha$ absorption in the enhanced densities at the boundaries of stellar astrospheres \citep{2004LRSP....1....2W}. This method has provided mass loss estimates for the winds of a sample of stars, and demonstrates a correlation between the mass loss rate per unit surface area and the X-ray surface flux. For the most active stars, however, these results suggest a possible decrease in the mass loss rate. The presence of this ``wind dividing line'' is potentially very important, particularly considering the effect that the winds of young and very active stars may have on the atmospheres of any orbiting exoplanets (e.g. \citet{2010Icar..210..539Z}). However, neither Zeeman-Doppler maps nor wind models of a handful of stars that span this diving line show any change in field geometry across it \citep{2016MNRAS.455L..52V,2017MNRAS.466.1542S}.

The nature of the winds of the most active stars that lie beyond the ``wind dividing line'' can only be determined by measuring mass loss rates in this regime, but this is the very region of parameter space where measurements are most sparse. The aim of this paper is to describe a new method for measuring wind mass loss rates that is uniquely suited to this regime. 

\section{Stellar prominences}
\begin{figure}
	\includegraphics[width=\columnwidth]{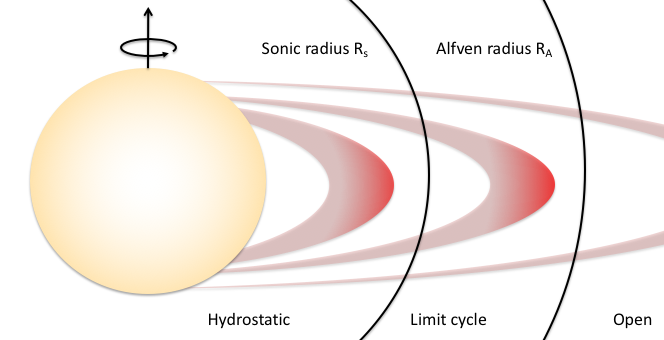}
    \caption{Schematic diagram showing prominence-bearing loops below the sonic radius (R$_s$) and between the sonic radius and the Alfv\'en radius (R$_A$). Beyond R$_A$, all field lines are open.}
    \label{cartoon}
\end{figure}
\begin{figure}
	\includegraphics[width=\columnwidth]{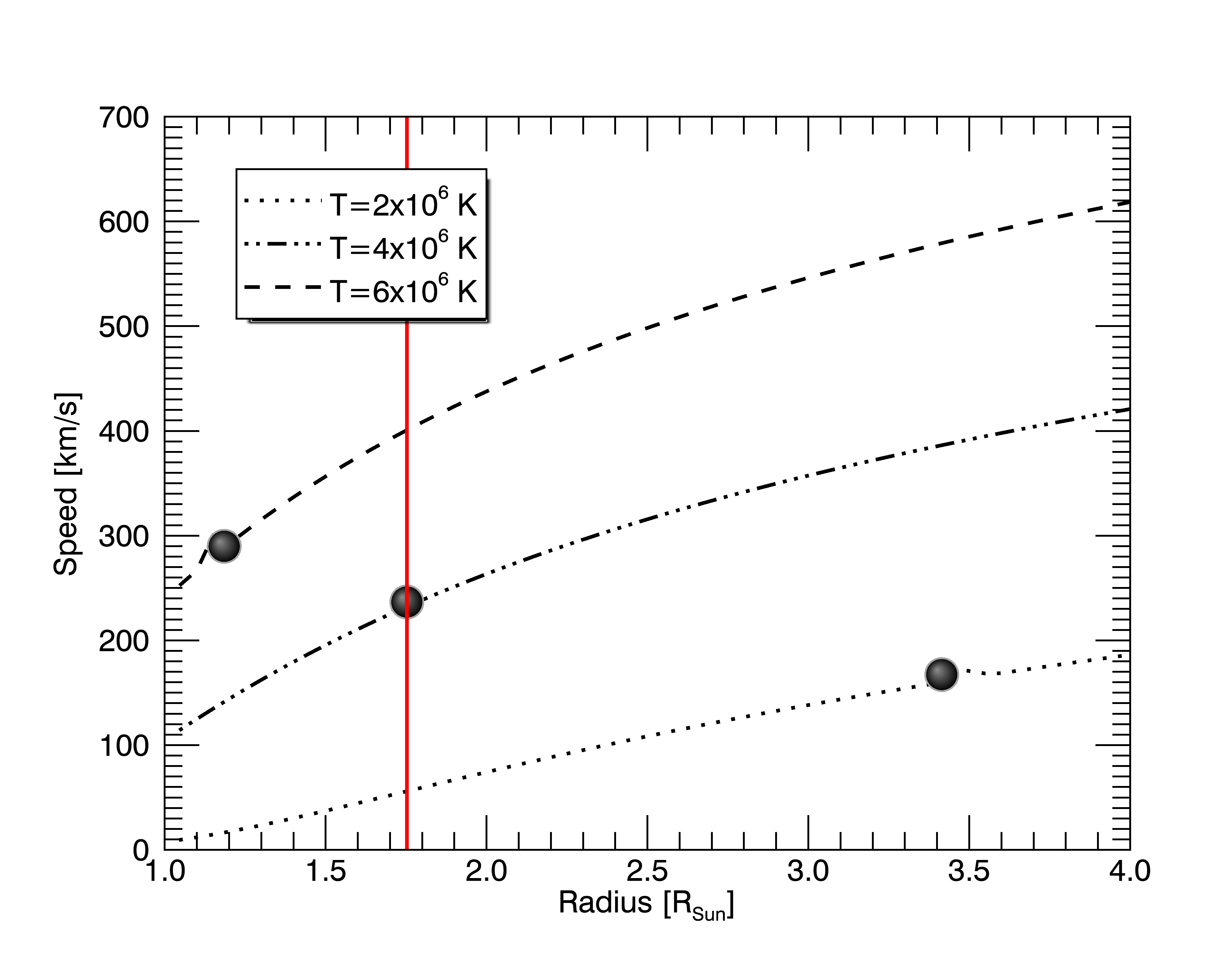}
    \caption{Upflow speed in a thermal wind at three temperatures. In each case the location of the sonic point is shown by a black dot. A red vertical line shows the location of the co-rotation radius for this 1M$_\odot$, 1R$_\odot$ star whose rotation period is 0.5 days. At the lowest of the three temperatures (dotted line), the wind accelerates slowly and the sonic point is above the co-rotation radius R$_k$. At the highest temperature (dashed line), the wind accelerates more quickly and reaches the sound speed below the co-rotation radius R$_k$. At the critical temperature (middle line), the sonic point coincides with the co-rotation radius R$_k$}
    \label{wind}
\end{figure}
The essential barrier to measuring cool star winds is the low density of the outflowing plasma, but some regions of higher density do exist within the coronae of active stars. These are the stellar ``slingshot'' prominences that were first observed as H$\alpha$ absorption transients on the young active dwarf AB Dor \citep{1986PASAu...6..308R, 1989MNRAS.236...57C}. The transients are caused by cool clouds of mainly neutral plasma trapped within the corona by the star's magnetic field. Follow-up observations by \cite{1989MNRAS.238..657C} revealed that the transients corotate with the star, recurring on the stellar rotation period, and have radial accelerations consistent with locations near the corotation radius. Their lifetimes were observed to be of order 2-3 days. \cite{1990MNRAS.247..415C} observed simultaneous  absorption transients in the Ca II H \&\ K and hydrogen Balmer lines, deriving mass estimates around 2 to $6\times 10^{17}$ g, some three orders of magnitude greater than those of quiescent solar prominences. Since their discovery, similar condensations have been observed in a range of stars, from spectral types G to M, in both single and binary stars and in main-sequence and pre-main sequence stars. Indeed the term ``slingshot prominence'' appears to have been coined by \cite{1996MNRAS.281..626S}, who observed co-rotating emission transients in low-excitation lines at equivalent locations during outbursts of the dwarf novae IP Peg and SS Cyg.

Typically, slingshot prominences are supported at or around the Keplerian co-rotation radius, which is the equatorial radius at which the effective gravitational acceleration experienced by a co-rotating particle is zero. This is a natural location at which to expect coronal material to collect. Beyond this radius, coronal plasma will be centrifugally driven outward, leading to over-dense loop summits \citep{1991SoPh..131..269J}. These high densities predispose the loop summits to thermal collapse. \citet{1988MNRAS.233..235C} solved for the thermal and mechanical equilibrium of such loops and found two classes of solutions - thermally stable hot loops, with temperature maxima at their summits, and thermally unstable cool solutions with temperature minima at their summits. Later models extending this to more general loop geometries and heating functions demonstrated that these cool solutions can be produced in a wide range of circumstances and can be identified with the observed prominences \citep{1997A&A...327..252F,1997A&A...321..177U}. 

Models of prominence support in both single and binary systems have since demonstrated that the observed  field strengths are sufficient to explain the prominence masses derived from the absorption transients \citep{1995A&A...298..172F,1996ApL&C..34..101J,1996A&A...305..265F,2000MNRAS.316..647F,2001MNRAS.324..201J}. These models assume, however, that the stellar coronal field is closed out to the distances of the observed prominence locations. This requires coronae that extend for many stellar radii in these rapid rotators. Confining hot coronal plasma out to many stellar radii is a significant challenge even for the high field strengths observed at the surface of these stars. This problem is removed, however, if the prominences are supported not within the X-ray bright corona, but at greater distance within the stellar wind. \citet{2005MNRAS.361.1173J} demonstrated that this is possible and that cool equilibria exist for loops that extend well out into the stellar wind, out to a maximum radius that is a simple function of the co-rotation radius. 

\section{Prominence formation}

The picture that emerges from these studies is that the formation of a condensation in the corona (perhaps due to a thermal instability) leads to a drop in pressure at the loop summit. Plasma will flow from the loop footpoints to re-establish pressure balance. The nature of this pressure-driven flow is the same as for a thermal wind. The flow accelerates with distance and reaches the sound speed at the sonic radius R$_s$. If the coronal condensation lies below this sonic radius, then the flow will be subsonic when it arrives (see Fig.~\ref{cartoon}). Pressure balance can be re-established on a sound travel time within the loop and the loop can adjust to a new equilibrium. In this case, we would expect to observe quasi-steady loops whose lifetimes are determined by timescales for evolution of the coronal field \citep{2014MNRAS.443.3251G,2016MNRAS.456.3624G}.

If, however, the coronal condensation lies above the sonic point, the upflow from the footpoints will be supersonic when it arrives. In this case, information cannot travel back to the surface to allow the upflow to adjust to the rising density in the loop summit and the density will continue to grow. At some stage the maximum density that can be supported by the field will be exceeded. At this point, the prominence mass will no longer be confined and will either fall back towards the surface if it is below the co-rotation radius, or be centrifugally ejected if it is above. In this case we expect to see repeated formation and ejection of prominences on timescales determined by the time taken for the upflow to supply the maximum density. \citet{2018MNRAS.475L..25V} have calculated this maximum density for a sample of stars for which field strengths are known and for which coronal temperatures can be estimated using the scaling with X-ray flux from \citet{2015A&A...578A.129J}. They conclude that the predicted masses and lifetimes of these prominences reproduce well the observed range of values.

A third possibility is that the coronal condensation forms above the Alfv\'en radius (at which the flow speed equals the Alfv\'en speed). In this case, the magnetic field cannot remain closed at this radius and we would not expect to detect any significant accumulation of material. For a low-$\beta$ plasma where the sound speed is less than the Alfv\'en speed, we expect the Alfv\'en radius to be greater than the sonic radius, as is shown in Fig. \ref{cartoon}.

We therefore expect three different types of behaviour, depending on where the prominence begins to form (see Fig.~\ref{cartoon}). Given that we observe prominences to form at or around the co-rotation radius R$_k$, the type of behaviour depends on the location of the co-rotation radius relative to the sonic and Alfv\'enic points. These three regimes can be expressed as:
\begin{itemize}
\item $R_k < R_s < R_A$  {\it Hydrostatic regime}: mechanical equilibria may be possible - the prominence lifetime is governed by the evolution of the coronal field.
\item $R_s < R_k < R_A$ {\it Limit cycle regime}: prominences form, grow and are ejected on a timescale determined by the accumulation time for the maximum mass that can be supported and
\item $R_s < R_A < R_k$ {\it Open field regime}: No closed loops exist to support prominences.
\end{itemize}

\section{The three regimes}

In order to explore which stars might exhibit these regimes, we need to determine the co-rotation and sonic radii. The equatorial co-rotation radius is defined as the location where the effective gravity $g_{\rm eff}$ is zero, such that if
\begin{equation}
g_{\rm eff} = -\frac{GM_\star}{R^2} + \Omega^2 R ,
\end{equation}
then  
\begin{equation}
R_k = \left( \frac{GM_\star}{\Omega^2} \right)^{1/3}
\end{equation}
where $\Omega$ is the stellar rotation rate. 
\begin{figure}
    	\includegraphics[width=\columnwidth]{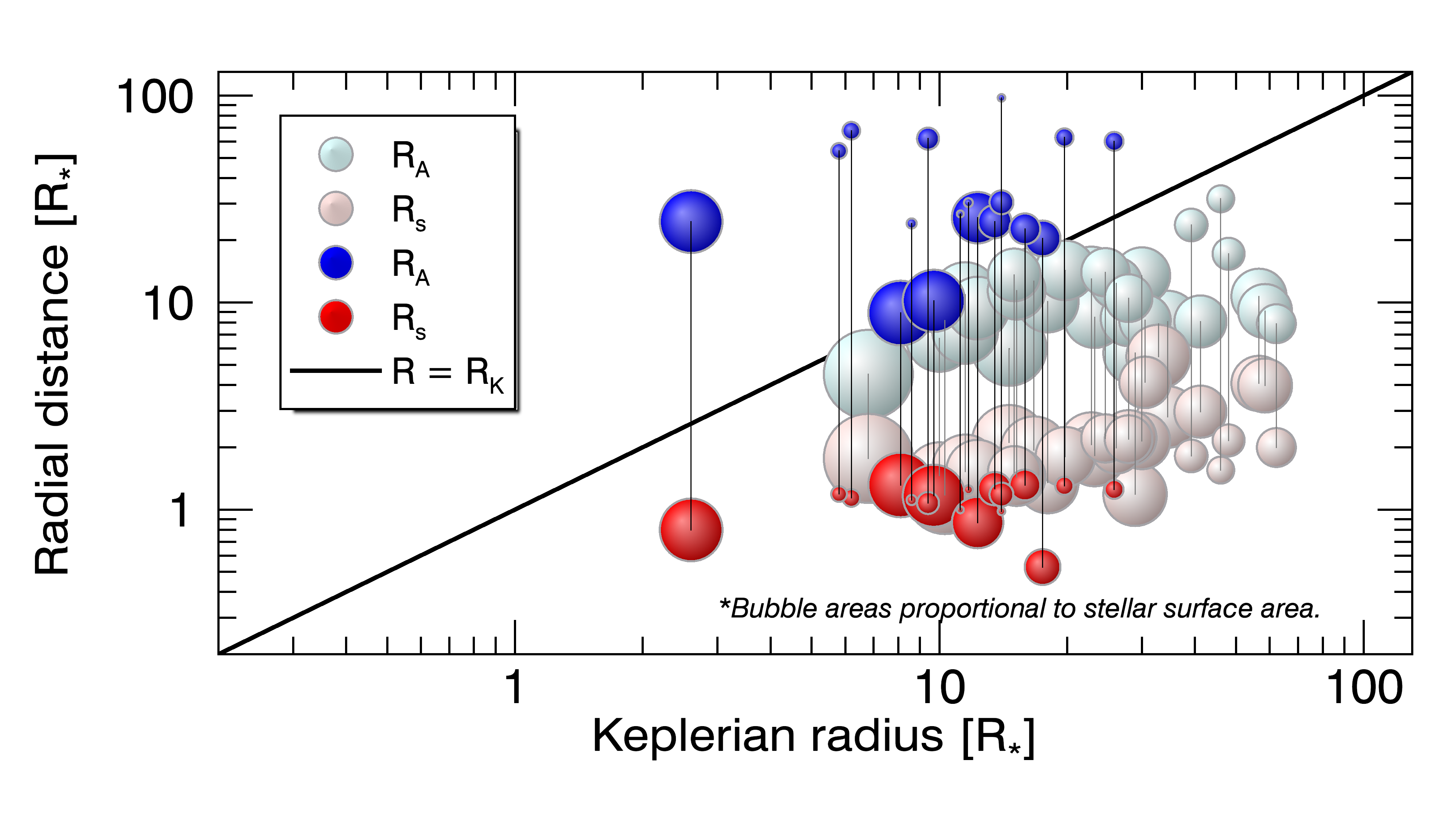}
    \caption{Alfv\'en (R$_A$) and sonic (R$_s$) radii for all the low mass stars in \citet{2018MNRAS.475L..25V}. Values for each star are connected by a black line. The co-rotation radius is shown as a diagonal thick black line. Faint symbols denote stars in the ``open field'' regime with $R_s < R_A < R_k$ where no equilibrium is possible. Darker symbols denote stars in the ``limit-cycle'' regime with $R_s < R_k < R_A$. There are no stars in this sample with $R_k < R_s < R_A$.}
    \label{RsRkRA}
\end{figure}

In order to determine the location of the sonic point, we consider a radial, isothermal, pressure-driven outflow \citep{1958ApJ...128..677P}. Mass conservation for a density $\rho$ and velocity $u$ gives
\begin{equation}
\dot{M} = 4\pi R^2\rho u
\end{equation}
and hence the momentum equation gives
\begin{equation}
\left( u^2-c_s^2 \right) \frac{d (\ln u)}{dR} = \frac{2c_s^2}{R} \left( 1-\frac{GM_\star}{2c^2_s R}\right)
\label{parker_1}
\end{equation}
where the sound speed is given by $c_s^2 = kT/m$ for a temperature $T$ and mean particle mass $m$. At the sonic point, $u=c_s$ and $R = R_s$ where $R_s$ is given by
\begin{equation}
R_s = \left( \frac{GM_\star}{2c_s^2} \right).
\label{sonic_radius}
\end{equation}
Equation (\ref{parker_1}) can be most conveniently written as
\begin{equation}
\left(\frac{u}{c_s}\right)^2 - \ln\left( \frac{u}{c_s}\right)^2 = 4\ln \frac{R}{R_s} + 4\frac{R_s}{R} -3.
\end{equation}
This equation has several well-known asymptotic forms \citep{1958ApJ...128..677P,1999isw..book.....L}, notably that close to the stellar surface, for  $R \ll R_s$ this reduces to
\begin{equation}
u(R)=c_s(R_s/R)^2e^{3/2}e^{-2(R_s/R)}.
\label{parker}
\end{equation}
In particular, this allows us to set $R=R_\star$ to recover the velocity ($u_\star$) close to the stellar surface and hence the mass loss rate
\begin{equation}
\dot{M} = 4\pi R_\star^2\rho_\star u_\star
\label{Mdot}
\end{equation}
where $\rho_\star$ is the mass density at the stellar surface.

Figure \ref{wind} illustrates the effect of the temperature on the wind speed. Hotter winds accelerate faster and reach the sonic point sooner. As a result, these may have a sonic point below the co-rotation radius. Loops at these temperatures therefore have R$_s$<R$_k$ and so lie in the ``limit-cycle'' regime. Much cooler winds accelerate more slowly and so may have a sonic point beyond the co-rotation radius. Loops at these temperatures therefore have R$_s$>R$_k$ and so lie in the ``hydrostatic'' regime. 

For each star (i.e. for each combination of M$_\star$, R$_\star$ and rotation period P) there is one temperature such that R$_s$=R$_k$. This is given by
\begin{equation}
T_{\rm crit} [10^6K] = 1.6 \left(\frac{M_\star [M_\odot]}{P [d]} \right)^{2/3}.
\label{Tcrit}
\end{equation}
The value of this critical temperature determines the fraction of the magnetic loops in the corona that lie in each regime. Loops cooler than this critical temperature will lie in the ``hydrostatic'' regime, while the hotter loops with summits beyond the sonic  point will lie in the ``limit cycle'' regime. As main-sequence stars age, we expect their rotation periods to lengthen as their winds carry away angular momentum. The critical temperature will therefore decrease with age. The overall level of magnetic activity of the star will also decrease as the star spins down. An average coronal temperature (weighted by the emission measure) can be determined from X-ray spectra. \citet{2015A&A...578A.129J} derived a scaling of $ T = 0.11F_X^{0.26}$ between this average coronal temperature and the X-ray flux. From this, they determined that in the unsaturated regime, $T \propto M_\star^{-0.42}/P^{0.52}$ while in the saturated regime, $T \propto M_\star^{0.6}$. Thus in the unsaturated regime, we have
\begin{equation}
\frac{T}{T_{\rm crit}} \propto \frac{P^{0.15}}{M_\star^{1.09}}
\end{equation}
while in the saturated regime
\begin{equation}
\frac{T}{T_{\rm crit}} \propto \frac{P^{0.67}}{M_\star^{0.06}}.
\end{equation}
In both cases, therefore, the critical temperature decreases faster with rotation period that the average coronal temperature. The more slowly-rotating stars may therefore support slingshot prominences over a greater range of temperatures than rapidly-rotating stars. 

Beyond a rotation period of a few days, however, the co-rotation radius has moved out beyond the Alfv\'en radius, the star is in the  ``open field'' regime and no slingshot prominences can be supported at all. Calculation of the Alfv\'en radius follows from a consideration of the balance of torques along the magnetic field. In the simplest {\it Weber-Davis} case of a radial magnetic field it can be expressed as
\begin{equation}
\frac{R_A}{R_\star} = \left( 1-\frac{R_\star B_R B_\phi}{\dot{M}\Omega}\right)^{1/2}
\end{equation}
where $B_R$ and $B_\phi$ are evaluated at the stellar surface \citep{1967ApJ...148..217W,1999isw..book.....L,2016MNRAS.458.1548B}. While the determination of the sonic point depends primarily on the temperature, this expression highlights the role of the strength and geometry of the stellar magnetic field in the determination of the Alfv\'en radius.

\section{Examples of the three regimes}
\begin{figure}
	\includegraphics[width=\columnwidth]{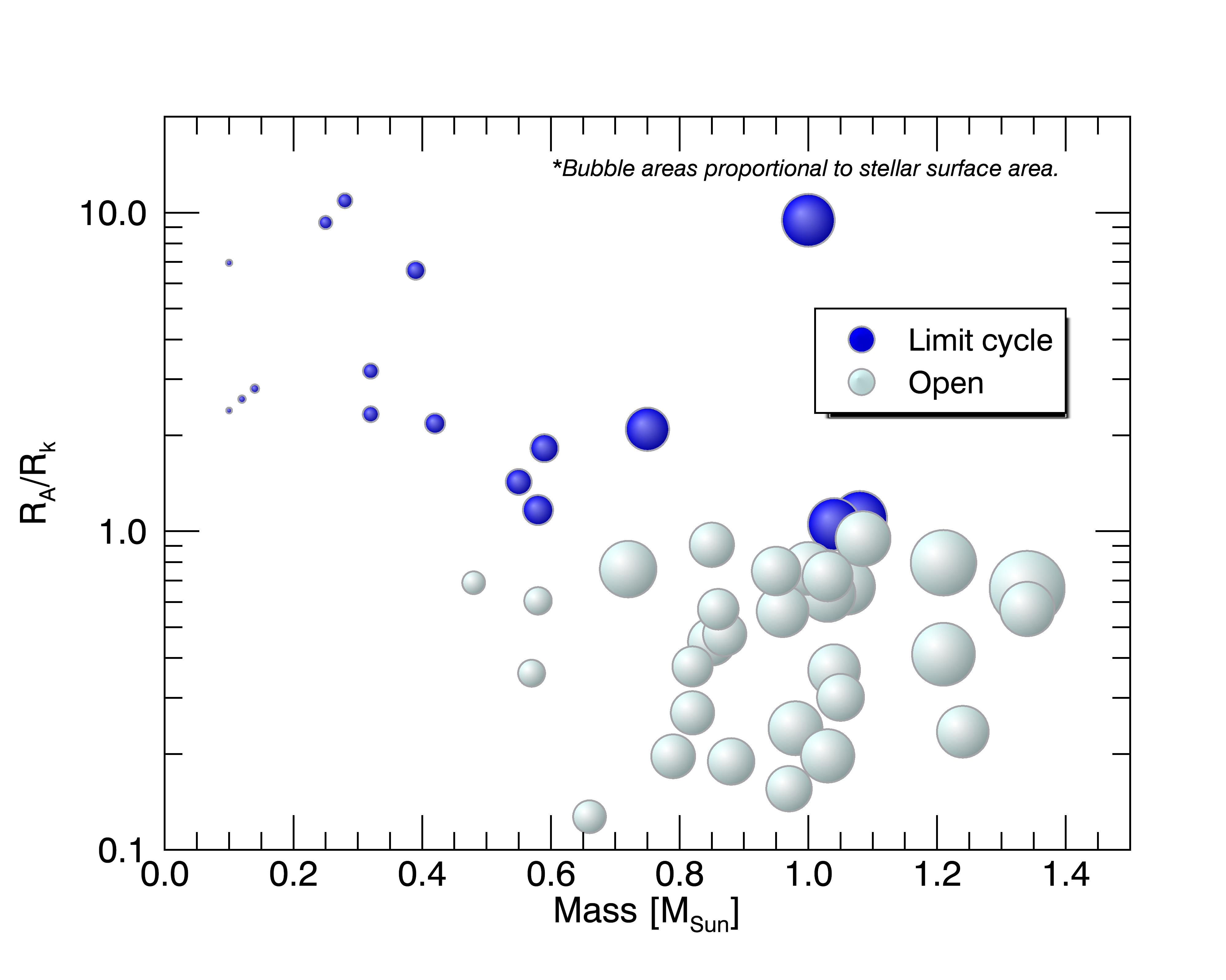}
    \caption{Ratio of Alfv\'en radius (R$_A$) to co-rotation radius (R$_k$) as a function of stellar mass. Stars in the limit cycle regime (those with R$_A$/R$_k \geq 1$) are shown as dark blue, while those in the open field regime (those with R$_A$/R$_k < 1$) are shown as light blue.}
    \label{RARk_mass}
\end{figure}
In order to quantify the distinction between the three regimes, we use the sample of low mass stars whose properties (including co-rotation radii) are tabulated in \citet{2018MNRAS.475L..25V}. For these stars, the Alfv\'en radius has been determined from a Weber-Davis wind solution, while the coronal temperature has been estimated from the X-ray flux using the relationship due to \citet{2015A&A...578A.129J}. This allows the sonic radius to be calculated using Eqn. (\ref{sonic_radius}).  

\begin{figure}
 	\includegraphics[width=\columnwidth]{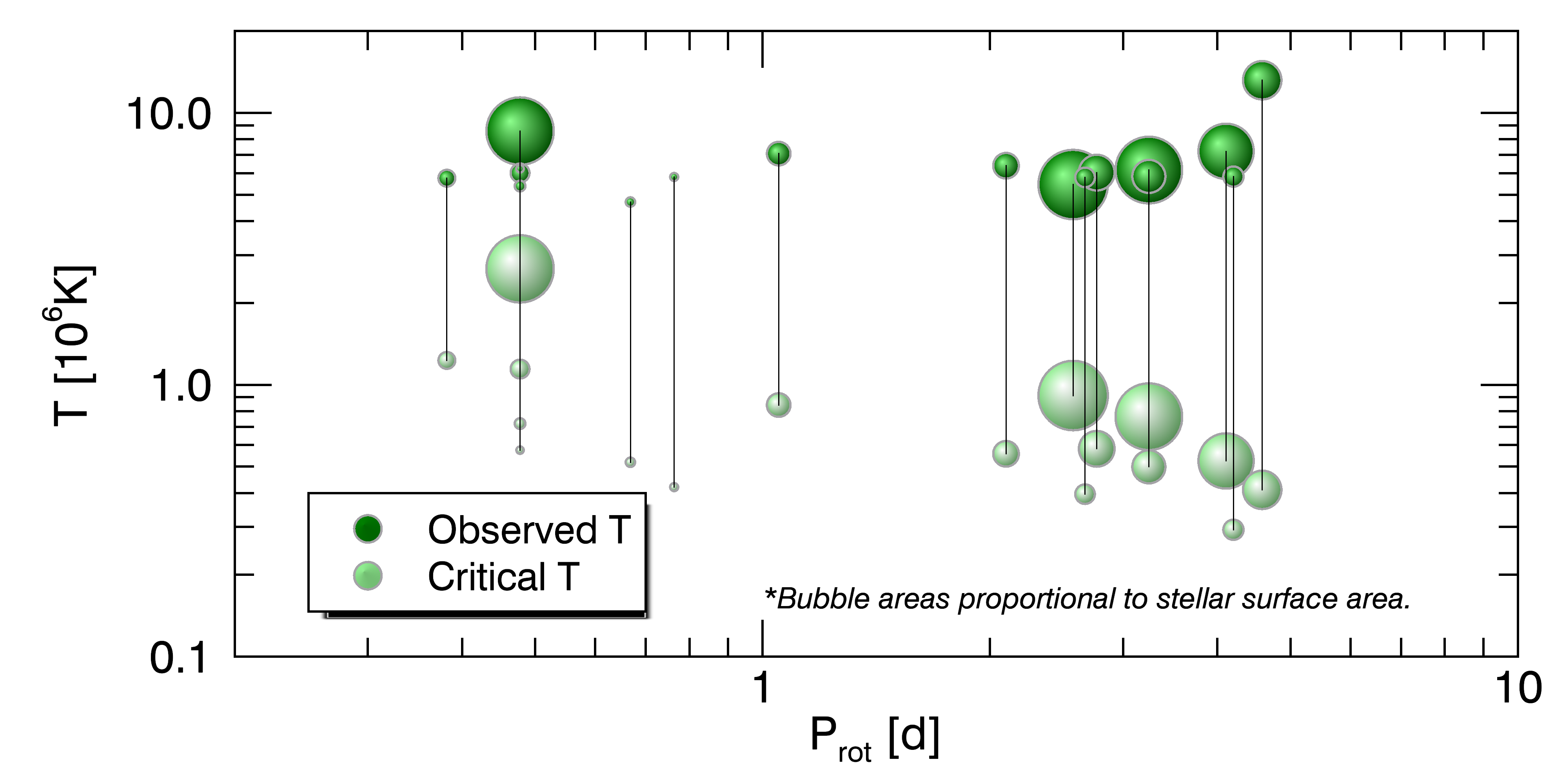}   
    \caption{Temperatures of stars that lie in the ``limit cycle'' regime, derived from their X-ray fluxes, using the relationship in \citet{2015A&A...578A.129J} (dark green circles). Also show in pale green are the minimum temperatures for the ``limit cycle'' behaviour. Any loops at temperatures below this may have cool summits that are in hydrostatic equilibrium.}
    \label{minT}
\end{figure}

Fig.~\ref{RsRkRA} shows the sonic, Alfv\'en and co-rotation radii for these stars. The faint symbols denote those stars that lie in the ``open field'' regime where $R_s < R_A < R_k$. We do not expect these stars to support significant numbers of slingshot prominences and so we do not consider them further. The darker symbols denote those stars that lie in the ``limit-cycle'' regime with $R_s < R_k < R_A$. We note that there are no stars in this sample that lie in the ``hydrostatic'' regime with $R_k < R_s < R_A$.

Although there is a trend for the more slowly-rotating stars to lie in the ``open field'' regime, there is some overlap in the two groups, principally due to the variation in the Alfv\'en radius. Notably, the stars in the ``limit-cycle'' regime tend to be of lower mass. This can be seen very clearly by plotting the ratio of the Alfv\'en radius to the co-rotation radius as a function of mass (see Fig. \ref{RARk_mass}). The lowest mass stars tend to have larger Alfv\'en radii, as a result of their larger dipole field strengths \citep{2017MNRAS.466.1542S}.  While few prominences have been detected in low mass stars, this is most likely to be due to their intrinsic faintness rather than the absence of prominences, which have generally been discovered as by-products of Doppler imaging studies which preferentially target bright stars. The clear outlier in this plot is the lone star in the top right-hand corner. This is AB Dor, which is a very rapid rotator with a strong field, and the first known example of the phenomenon.

In order to illustrate the range of coronal temperatures that lie in the limit cycle regime, we show in Fig~\ref{minT} both the observationally-derived temperature for each of the limit-cycle stars in Fig~\ref{RsRkRA} and also the critical temperature below which hydrostatic solutions are possible. This critical temperature is significantly below the X-ray determined temperature for all these stars, but still within the range of coronal temperatures that characterise stellar coronae. 
Within the corona of each star we may expect a range of loop temperatures, reflecting the different mechanisms responsible for their formation and heating. For example, loops in active regions may be associated with flaring and filled with plasma evaporated from the chromosphere, while loops reforming after a mass ejection may be heated by the reconnection process directly. Loops whose footpoints are in quiet areas of the surface may be cooler than either of these cases. Fig.~\ref{minT} shows that the range of loop temperatures over which limit-cycle behaviour may be expected is quite large, suggesting that the repeated formation and ejection of prominences may be common. Prominences may also form of course in loops below the critical temperature, but they will lie in the hydrostatic regime. We note also that, as expected from Eqn. (\ref{Tcrit}), this critical temperature is a decreasing function of rotation period. As the co-rotation radius moves outward with increasing rotation period, so loops need to be progressively cooler in order that their sonic point coincides with the co-rotation radius.  

\section{Prominences as wind gauges}
From the sample of stars in \citet{2018MNRAS.475L..25V} we therefore deduce that a significant number of the lowest mass stars are likely to exhibit repeated formation and ejection of slingshot prominences. For these stars, the prominence formation sites act as temporary wind traps, storing wind material in cool condensations whose mass grows until it can no longer be supported. These condensations of wind material become observable when their mass provides sufficient optical depth that they can be detected in H$_\alpha$ as absorption transients. 
\begin{figure}
\begin{centering}
 	\includegraphics[width=5cm]{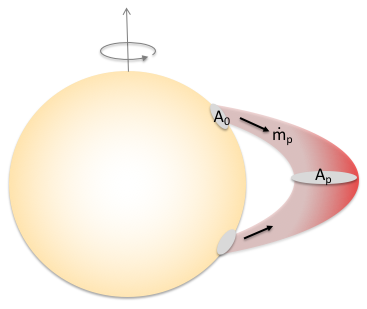}   
    \caption{Schematic view of the loop geometry. Mass flows from the loop footpoints at a rate $\dot{m}_p$. The cross-sectional loop area is A$_0$ at the surface and A$_p$ at the prominence location.}
\end{centering}
    \label{upflow_cartoon}
\end{figure}
\begin{table*}
	\centering
	\caption{Data for stars with published prominence masses. Columns show, respectively, the stellar name, mass, radius, rotation period, prominence mass, prominence lifetime, X-ray flux, calculated coronal temperature, critical coronal temperature for limit cycle behaviour and the predicted wind mass loss rate. Optical data are taken from \citet{2000MNRAS.316..699D,1990MNRAS.247..415C,2006MNRAS.373.1308D,2016A&A...590A..11V,1990ApJ...349..608Y,1996A&A...311..651B,2001MNRAS.326..950B,2016MNRAS.463..965L} while X-ray data are taken from  \citet{2002AJ....124.1670M,2013A&A...560A..69L,2009A&A...499..129L,2003A&A...397..147P}.}
	\label{tab:example_table}
	\begin{tabular}{lccccc|ccc|r} 
		\hline
		Star & Mass  & Radius &  P & m$_p$ & $\tau$ & F$_X$ & T$_{\rm cor}$ & T$_{\rm crit}$ & $\dot{\rm{M}}_\star$ \\
           & [M$_\odot$] & [R$_\odot$] & [d] & [10$^{15}$ g] & [d] & [10$^7$ erg cm$^{-2}$s$^{-1}$] & [10$^6$ K] & [10$^6$ K] & [10$^{-14}$ M$_\odot$/yr] \\
		\hline
		LQ Lup & 1.16 & 1.3 &  0.31 & 20000 & <4 & 7.8 & 12.4 & 3.9 & 9000\\
		AB Dor          & 1.0 & 0.93 &  0.514 & 200 - 600 & 1 & 1.4 & 7.9 & 2.5 & 700\\
		Speedy Mic      & 0.82 & 1.06 &  0.38 & 50 - 230 & 1 & 17.0 & 15.2 & 2.7 & 260\\
        V374 Peg        & 0.3 & 0.34 &  0.44570 & 10 & 1/15 - 1/60 & 0.4 & 5.7 & 1.2 & 400\\
        HK Aqr          & 0.4 & 0.59 & 0.43 &  57 & 1 & 0.48 & 6.0 & 1.5 & 100\\
		\hline
	\end{tabular}
    \label{data}
\end{table*}

We can use the {\it observed} masses and lifetimes to infer the mass upflow rate into the prominences, since 
\begin{equation}
\dot{m}_p \sim \frac{m_p}{\tau}
\end{equation}
where $m_p$ is the observed prominence mass, $\tau$ is the lifetime and $\dot{m}_p$ is the rate at which mass flows through the two footpoints of the prominence-bearing loop (see Fig. \ref{upflow_cartoon}). If we know the surface area of each loop footpoint ($A_0$), we can determine the mass flow rate per unit surface area of the star and hence the mass loss rate in the stellar wind from 
\begin{equation}
\dot{M}_\star = 4\pi R_\star^2\frac{\dot{m}_p}{2A_0}
\end{equation}
where we have assumed that both loop footpoints contribute to the mass flux into the prominence. 

Estimating the loop footpoint area is straightforward if the area $A_p$ of the prominence is known. Flux conservation ensures that along the prominence bearing loop, $B*A$ is conserved. If we can estimate the field geometry, then a measure of the prominence area in the corona gives the footpoint area $A_0$ directly. At the heights at which these prominences are observed, the dipole component of the magnetic field dominates, so we may assume that 
\begin{equation}
A_0 = A_p \left( \frac{R_\star}{R_p} \right)^3
\end{equation}
where $R_p$ is the radial height of the prominence. Several estimates of prominence areas exist. \citet{1990MNRAS.247..415C} estimate that prominences on AB Dor occult 20$\%$ of the stellar disk at a distance of 2.7R$_\star$ from the stellar rotation axis, while for LQ Lup, \citet{2000MNRAS.316..699D} estimate that the prominences have a linear extent of 0.7R$_\star$ at a distance of 1.65$R_\star$. In a similar way, \citet{2016MNRAS.463..965L} determine a prominence area of 4$\%$ of the stellar surface area. This gives values for the area of each footpoint of 0.3$\%$, 0.7$\%$ and 0.1$\%$ of the stellar surface area respectively, such that $\dot{M}_\star = 197\dot{m}_p$ (AB Dor), $\dot{M}_\star = 73\dot{m}_p$ (LQ Lup) and $\dot{M}_\star = 415\dot{m}_p$ (HK Aqr). 

\begin{figure}
	\includegraphics[width=\columnwidth]{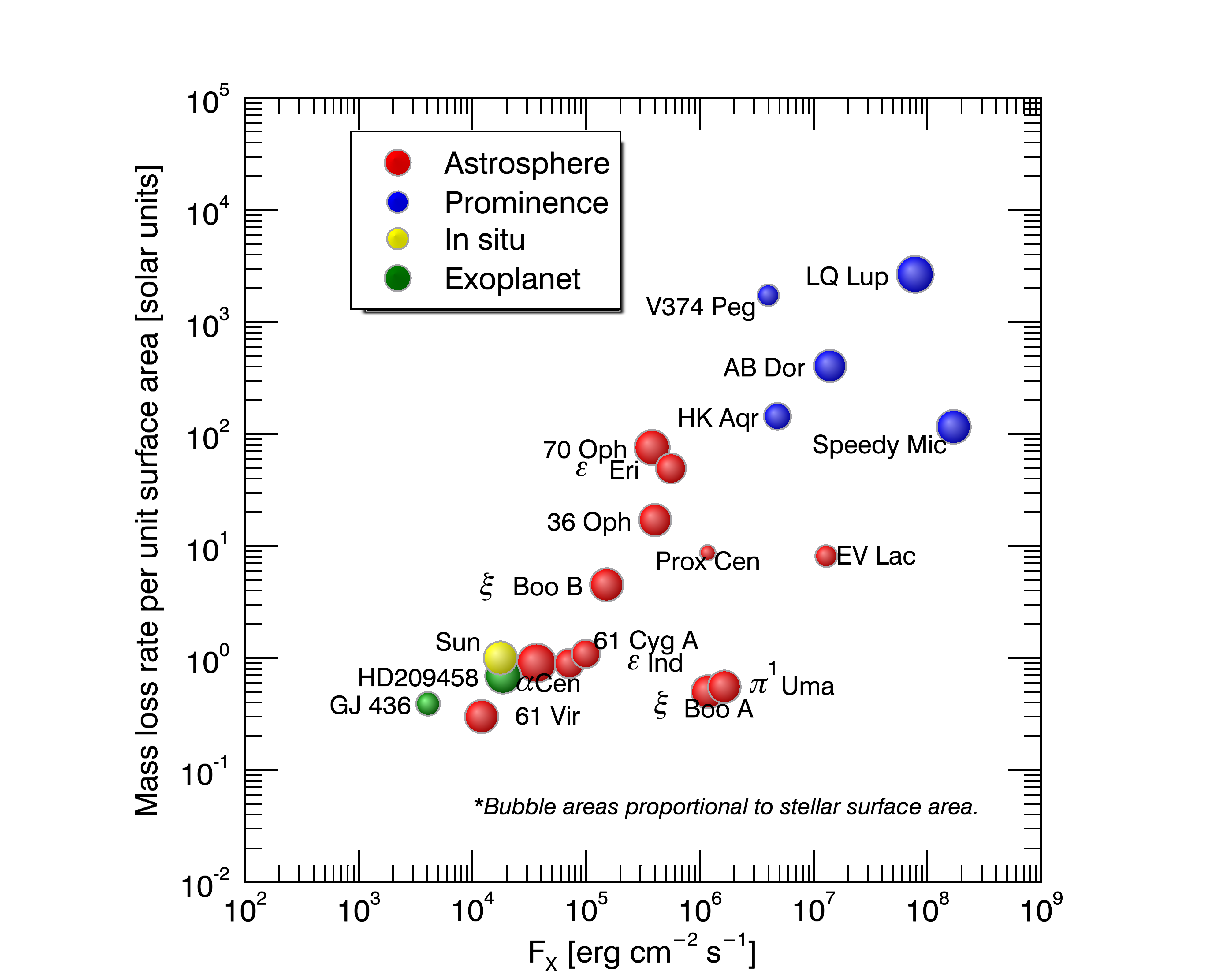}
    \caption{Mass loss rate in the stellar wind (per unit stellar surface area) plotted against the X-ray surface flux. Results are shown for estimates of mass loss rates made by a variety of methods \citep{2005ApJ...628L.143W,2010ApJ...715L.121W,2017MNRAS.470.4026V,2014Sci...346..981K}. The blue symbols represents those stars for which the prominence recurrence times are used.}
    \label{wood}
\end{figure}

Guided by these values, we assume that the two footpoints of prominence bearing loops typically cover 1$\%$ of the stellar surface, such that
\begin{equation}
\dot{M}_\star \sim 100 \frac{m_p}{\tau}.
\end{equation}
Table~\ref{data} shows parameters and predicted wind mass loss rates for the 5 stars for which prominence masses and lifetimes have been measured: LQ Lup, AB Dor, Speedy Mic, V374 Peg and HK Aqr. In all five cases the coronal temperature derived from the X-ray flux is significantly higher than the critical temperature. For these stars, therefore, the sonic point is above the co-rotation radius and the stars lie in the limit-cycle regime. The resulting wind mass loss rates per unit surface area are shown in Fig.~\ref{wood}, beside the other values derived from a range of methods.

\section{Discussion}
We have considered the nature of the upflows required to form prominences at the co-rotation radius of rapidly-rotating stars. We find three regimes, depending on the relative locations of the sonic point, the co-rotation radius and the Alfv\'en radius. In two of these regimes, prominences may be supported within the star's magnetic field, either in a static form if the upflow is subsonic at the prominence formation site, or in a form of a ``limit cycle'' if the upflow is supersonic here. 

The principal difference between these two regimes is that in the limit-cycle case, the upflow from the stellar surface along the prominence-bearing loop becomes supersonic before it reaches the prominence. The surface therefore cannot adjust to the formation of the prominence and so the mass flow is similar to that in the open wind. This upflow is effectively continuous, even although the release of this material when the prominence becomes unstable is quasi-periodic. The result of this is to convert field lines that might otherwise have been closed into wind-bearing field lines. In the case of the hydrostatic regime, the wind material may be stored from some time in a stable prominence and the surface upflow will cease. Although these field lines also contribute to the wind, they do so in an occasional, rather than a quasi-periodic limit-cycle fashion. This behaviour shares some common features with the slow component of the solar wind, which is released from above active regions and has a characteristic speed that is lower than that of the fast component \citep{2010ApJ...715L.121W}.

In both cases, the release of prominence material carries away a specific angular momentum $L_{\rm prom}=R_k^2 \Omega$ compared to the value $L_{\rm wind}=R_A^2 \Omega$ carried by the stellar wind. Since for these stars, $R_k < R_A$, even if the prominence-bearing loops covered the same fraction of the stellar surface as the wind-bearing field lines, they would remove less angular momentum.

There are two underlying assumptions of our model that may affect the wind mass loss rates that we predict. The first is that we assume that the winds and coronae of these stars are at the same temperature. The wind upflow that forms the prominences is therefore at the same temperature as the ambient stellar wind that flows along the open field lines. We expect however, that a range of temperatures will exist within the coronae and winds of these stars, and indeed Fig.~\ref{minT} shows that limit-cycle behaviour can be expected for a wide range of temperatures. If the stellar wind were, on average, significantly cooler than the prominence-bearing loops, then we would over-estimate the wind mass loss rate. We can quantify this using the approximation for the wind speed given by (\ref{parker}), or directly from Fig~\ref{wind}. If we assume that the base pressures are constant, then the base density scales as $1/T$, so a cooler wind is more dense. The velocity of a cooler wind is lower, however, and this effect tends to dominate the mass flow rate $(\rho u$) per unit surface area. Using the example in Fig.~\ref{wind}, we can see that the ratio of mass fluxes for winds at temperatures of $2\times 10^6$K, $4\times 10^6$K, and $6\times 10^6$K would be in the ratio 1:6:9. We note however that it is unlikely that the winds of the active stars in which slingshot prominences are observed would be cooler than the wind of the relatively inactive Sun. Estimates of the temperatures at the base of the fast and slow solar wind are $3.8\times 10^6$K and $1.8\times 10^6$K respectively \citep{2015A&A...577A..27J}. This is above the critical temperature for most of the stars in our sample, suggesting that few, if any, would lie in the hydrostatic regime.

The second assumption is that the prominence-bearing loops have a dipole geometry and hence, since the magnetic flux is conserved, the loop areas at the prominence location and at the surface vary inversely as the cube of the radius. If the expansion of the loops is less than this (perhaps because they have a degree of internal twist providing extra support against expansion) then our method will underestimate the loop footpoint areas and hence overestimate the corresponding wind mass loss rates. The most extreme case of this would be for a loop of constant cross-section where the loop footpoint area is the same as the prominence area. Recalculating our wind mass loss rates in this extreme case would give $\dot{M}_\star = 10\dot{m}_p$ (AB Dor), $\dot{M}_\star = 17\dot{m}_p$ (LQ Lup) and $\dot{M}_\star = 13\dot{m}_p$ (HK Aqr). In this case, our wind mass loss rates would be at most an order of magnitude lower. 

There are also two aspect of the observations that may lead us to underestimate the wind mass loss rates. The first is that the mass we observe is in most cases not the total mass of the prominence. Only the part of the prominence that transits the stellar disk can be observed in absorption, so the masses determined observationally may underestimate the prominence mass and hence underestimate the wind mass loss rate. It is notable that the only star in our sample that is viewed almost pole-on, and for which therefore all the prominences are in view, is LQ Lup, which also has the largest prominence mass and the largest mass loss rate.

The second aspect of the observations that may lead to an underestimated mass loss rate concerns the prominences in the static regime. These may be destabilised by processes such as field evolution due to surface differential rotation, meridional flow or flux emergence. Whatever the cause of destabilisation, however, the mass that can be accumulated on the observed lifetime is determined by the upflow rate of plasma that forms the prominence. If a prominence lies in the stable regime, and its formation time is short compared to the interval between observations, then its mass may have been constant for some fraction of the lifetime determined by observations. In this case, our method would underestimate the mass flow rate by overestimating the time taken for the mass to accumulate.

The values for the wind mass loss rate shown in Fig. \ref{RsRkRA} are also likely to show a large scatter because they are derived from observations of a small number of prominences at single epochs, and so are subject to Poisson statistics. We do not know the distribution of prominence masses and lifetimes on many of these stars.

We note also that the prominence-bearing stars in Fig. \ref{wood} span a range of ages (both pre-main sequence and main-sequence) and also stellar masses and hence internal structures. M dwarfs are particularly likely to host slingshot prominences because of their strong, dipolar fields and their rapid rotation rates \citep{2008MNRAS.390..545D,2008MNRAS.390..567M,2010MNRAS.407.2269M}. For pre-main sequence stars also, the strong field \citep{2016MNRAS.457..580F,2018MNRAS.474.4956F} makes them likely candidates. Their inflated radii give larger pressure scale heights but smaller ratios of the co-rotation radius to the stellar radius. The co-rotation radius is therefore more likely to lie within the closed field corona where prominences may form. The detection of slingshot prominences in several weak-lined T Tauri stars \citep{2008MNRAS.385..708S,2009MNRAS.399.1829S} therefore not only provides information about the structure of the coronae of these very young stars, but also the nature of their mass loss rates. 

The formation and centrifugal support of slingshot prominences is a feature of rapid rotation and therefore high X-ray flux. This makes these prominences ideal tracers of mass loss in the most active stars where other mass loss estimates are currently lacking. Extrapolating from the mass loss estimates of less active stars, \citet{2005ApJ...628L.143W} suggested that the more powerful wind of the younger, more active Sun could have eroded the atmospheres of solar system planets  and in particular contributed to the loss of the Martian atmosphere. Astrospheric estimates of mass loss rates in stars with higher levels of X-ray flux appear to show a decline beyond a critical ``wind dividing line'', however. One possible explanation, proposed by \citet{2005ApJ...628L.143W}, is that the polar spots often observed in very active stars (see \citet{2009A&ARv..17..251S}) indicate a strong dipolar field that could suppress the action of a wind. The strong toroidal field often observed in very active stars \citep{2015MNRAS.453.4301S} has also been proposed as mechanism to choke the stellar winds of these stars \citep{2010ApJ...717.1279W}. In a study of the magnetic field topologies of stars on either side of the ``wind dividing line'', \citet{2016MNRAS.455L..52V}, however, concluded that no significant transition was apparent at this boundary.

Our wind mass loss estimates suggest that the original increase in $\dot{\rm M}$ with F$_X$ determined by \citep{2002ApJ...574..412W} continues to the highest X-ray fluxes observed, albeit with perhaps a large scatter in values. At present, the observations of prominences are biased towards the ultra-fast rotators, since it is in these systems that the co-rotation radius is close to the star and so the prominences are most likely to occult the stellar disk. Modelling of rotational evolution of solar analogues with distributions of initial rotation rates, (eg  \citet{2015A&A...577A..27J,2015A&A...577A..28J}) supports the idea that these stars may be particularly rapid rotators. Indeed, at these high wind mass loss rates, the upper limits provided by radio observations, (eg \citet{2017A&A...599A.127F}) can provide important constraints.  The prominence-bearing stars in this high-activity regime have a variety of field topologies, demonstrating that prominence support is a common feature, whenever the star is rotating sufficiently rapidly that the co-rotation radius comes inside the Alfv\'en radius. 

The observation of significant mass loss in slingshot prominences raises the question of the mass loss in any associated Coronal Mass Ejections (CMEs). The relationship between prominences and CMEs on stars is not well understood, and not currently constrained observationally. We note that we use the term ``prominence'' to refer to cool material in the stellar corona, rather than the hot material associated with solar CMEs. The intriguing possibility that in young stars, the mass loss from CMEs might contribute significantly to (or even dominate over) the stellar wind has been suggested \citep{2013AN....334...77A,2017ApJ...840..114C} but awaits more observations of stellar CMEs \citep{2014MNRAS.443..898L,2017MNRAS.472..876O} for confirmation.

Extrapolating from the relationship between the energies in solar flares and CMEs suggests that on the most active stars, CME ejection must be suppressed, to avoid an unphysically large energy flux in stellar CMEs \citep{2013ApJ...764..170D}. Further study of stellar slingshot prominences may clarify the geometry of the coronal magnetic field structures that confine them and hence shed some light on the nature of any CMEs that might be ejected with them.

\section{Summary and Conclusions}
Rapidly rotating low mass stars often display repeated formation and ejection of slingshot prominences. They typically, also, have very hot coronae.  The upflows that feed these prominences therefore become supersonic before they reach the prominence formation site and so the stellar surface cannot adjust to the presence of the prominence. It supplies mass to the prominence at the same rate as it supplies mass to the stellar wind. The prominence formation sites therefore act as ``wind gauges'', storing wind material until its density is high enough that it can be detected in H$_\alpha$ absorption. As a result, the observed recurrence times and masses of these prominences can be used to estimate the wind mass loss rates, as shown in Fig~\ref{wood}. This extends to 5 orders of magnitude the range of observed mass loss rates for low mass stars. 

This technique is only possible in the regime of rapid rotation, but this is the very regime in which mass loss estimates are most needed. The observations of prominences in stars with high X-ray fluxes shows no evidence of the ``wind-dividing line'', beyond which a suppression of mass loss rates of stars was suspected. Mass loss rate measurements based on prominence recurrence rates suggest that wind mass loss rates continue to increase with increasing X-ray flux. 

\section*{Acknowledgements}

We thank the referee for helpful feedback and acknowledge funding from STFC consolidated grant ST/R000824/1.




\bibliographystyle{mnras}
\bibliography{prom2018} 

\bsp	
\label{lastpage}
\end{document}